\documentclass[pra,aps,amsmath,amssymb,showkeys,eqsecnum]{revtex4}
\usepackage{graphicx}
\newcommand{\hh}{\mathcal{H}}

\newcommand{\clx}{{\mathcal{X}}}
\newcommand{\clz}{{\mathcal{Z}}}
\newcommand{\clf}{{\mathcal{F}}}
\newcommand{\clm}{{\mathcal{M}}}
\newcommand{\zset}{\mathbb{C}}
\newcommand{\mset}{\mathbb{M}}
\newcommand{\iu}{{\mathtt{i}}}
\newcommand{\am}{{\mathsf{A}}}
\newcommand{\mn}{{\mathsf{M}}}
\newcommand{\um}{{\mathsf{U}}}

\newcommand{\wm}{{\mathsf{W}}}

\newcommand{\bro}{{\boldsymbol{\rho}}}
\newcommand{\vbro}{{\boldsymbol{\varrho}}}
\newcommand{\vark}{\varkappa}
\newcommand{\bigma}{\bar{\sigma}}
\newcommand{\Tr}{{\mathrm{Tr}}}
\newcommand{\id}{{\mathrm{id}}}
\begin{document}
\clearpage
\preprint{}

\title{Flavor-mass majorization uncertainty relations and their links to the mixing matrix}

\author{Alexey E. Rastegin, Anzhelika M. Shemet}
\affiliation{Department of Theoretical Physics, Irkutsk State University, 664003, Russia}

\begin{abstract}
Uncertainties in flavor and mass eigenstates of neutrinos are
considered within the majorization approach. Nontrivial bounds
reflect the fact that neutrinos cannot be simultaneously in flavor
and mass eigenstates. As quantitative measures of uncertainties, both
the R\'{e}nyi and Tsallis entropies are utilized. Within the current
amount of experience concerning the mixing matrix, majorization
uncertainty relations need to put values of only two parameters,
viz. $\theta_{12}$ and $\theta_{13}$. That is, the majorization
approach is applicable within the same framework as the
Maassen--Uffink relation recently utilized in this context. We also
consider the case of detection inefficiencies, since it can
naturally be incorporated into the entropic framework. Short
comments on applications of entropic uncertainty relations with
quantum memory are given.
\end{abstract}


\maketitle

\pagenumbering{arabic}
\setcounter{page}{1}

\section{Introduction}\label{sec1}

The Heisenberg uncertainty principle \cite{heisenberg} is one of
cornerstones of modern physics. The traditional position-momentum
uncertainty relation was formally derived by Kennard \cite{kennard}.
For any pair of observables, the corresponding formulation was
presented by Robertson \cite{robert}. This formulation was
criticized for several reasons \cite{deutsch,maass}. Entropic
uncertainty relations were proposed as an alternative to more
traditional approach dealing with the product of standard
deviations. For the case of canonically conjugate variables, this
approach was initiated by Hirschman \cite{hirs} and later developed
in \cite{beck,birula1}. In effect, it seems to be more
fruitful for discrete observables. For basic advantages of the
entropic approach, see \cite{ww10,brud11,cbtw17,cerf19}.
Another reason to use entropic uncertainty relations is inspired by
the role of ``side information'' \cite{BCCRR10}.

Entropic uncertainty relations of the Maassen--Uffink type
\cite{maass} are actually Kraus' conjecture \cite{kraus87} proved on
the base of Riesz's theorem \cite{riesz27}. There are efforts to
formulate entropic uncertainty relations beyond this restriction.
Sometimes, the corresponding optimization can be carried out
explicitly \cite{rans12}. The authors of \cite{zbp13,zbp14}
examined a constrained optimization problem without the conjugacy
restriction on the entropic indices. Majorization relations are a
way to characterize uncertainties in terms of probabilities {\sl per
se} \cite{prtv11}. First majorization entropic uncertainty relations
were based on tensor products of probability vectors
\cite{prz13,fgg13}. Stronger bounds obtained in \cite{rpz14}
are based on majorization relations applied to direct sums of
probability vectors.

New interest to uncertainty relations is stimulated by recent
progress in quantum information science \cite{ww10,cbtw17}. In the
context of particle physics, uncertainty relations have found less
attention than they deserve. Neutrino physics had come across
radical ideas right since its appearing. One of explanations of
experimental data in studying $\beta$-decay claimed that
conservation laws hold only statistically \cite{zuber12}. Pauli's
solution was an alternative to such proposals. Neutrino oscillations
predicted by Pontecorvo \cite{bruno57} were for a time the
long-standing question \cite{kajita16,mcdd16}. The flavor
eigenstates of neutrinos and the mass ones form two different bases
related by the Pontecorvo--Maki--Nakagawa--Sakata (PMNS) matrix
\cite{bruno58,mns62}. Reference \cite{schwindt20} studied
corollaries of this fact via both the Robertson and Maassen--Uffink
formulations. There are physically realizable situations, where the
above issues are applicable.

The following issues were discussed as an arena for the use of
flavor-mass uncertainty relations \cite{schwindt20}. Studies of
cosmic neutrinos inevitably involve processing of large amounts of
data. Entropic functions are a standard tool in such questions. Being
almost free from decoherence effects, neutrinos are candidates to
implement some protocols of quantum information. Entangled states of
neutrino-lepton pairs produced by decaying gauge bosons can be used
to probe experimentally parameters of the mixing matrix. This work
is devoted to flavor-mass majorization uncertainty relations. In
Section \ref{sec2}, we briefly recall majorization uncertainty
relations and related material, including results of 
\cite{cp2014}. In Section \ref{sec3}, majorization approach is
applied to the flavor and mass eigenbases. The case of detection
inefficiencies is addressed as well. In Section \ref{sec4}, entropic
uncertainty relations for entangled neutrino-lepton states are
shortly discussed. In Section \ref{sec5}, we conclude the paper.

\section{On majorization uncertainty relations in general}\label{sec2}

For two integers $m,n\geq1$ the symbol $\mset_{m\times{n}}(\zset)$
denotes the space of all $m\times{n}$ complex matrices. For any
$\am\in\mset_{m\times{n}}(\zset)$, the square matrices
$\am^{\dagger}\am$ and $\am\am^{\dagger}$ have the same non-zero
eigenvalues. Taking the square root of these eigenvalues, we obtain
non-zero singular values $\sigma_{j}(\am)$ of $\am$ \cite{hj1990}.
We further use the spectral norm of $\am$ defined as
\begin{equation}
\|\am\|_{\infty}=\underset{j}{\max}\, \sigma_{j}(\am)
\, .
\label{spnm}
\end{equation}
Let $\clx=\bigl\{|x_{i}\rangle\bigr\}$ and
$\clz=\bigl\{|z_{j}\rangle\bigr\}$ with $i,j=1,\ldots,d$ be two
orthonormal bases. For the pre-measurement state $\bro$, the
probability vectors $p$ and $q$ have elements
$p_{i}=\langle{x}_{i}|\bro|x_{i}\rangle$ and
$q_{j}=\langle{z}_{j}|\bro|z_{j}\rangle$.

For the given probability distribution, the Shannon entropy is
written as
\begin{equation}
H_{1}(p)=-\sum\nolimits_{i} p_{i}\ln{p}_{i}
\, . \label{shent}
\end{equation}
Let $\eta_{1}$ be the largest among moduli
$\bigl|\langle{x}_{i}|z_{j}\rangle\bigr|$. Maassen and Uffink proved
that \cite{maass}
\begin{equation}
H_{1}(\clx;\bro)+H_{1}(\clz;\bro)\geq-2\ln\eta_{1}
\, , \label{unrshn}
\end{equation}
where $H_{1}(\clx;\bro)$ and $H_{1}(\clz;\bro)$ are obtained by
substituting the above probabilities. Let $\eta_{2}$ denote the
second largest value among
$\bigl|\langle{x}_{i}|z_{j}\rangle\bigr|$; then \cite{cp2014}
\begin{equation}
H_{1}(\clx;\bro)+H_{1}(\clz;\bro)\geq-2\ln\eta_{1}+(1-\eta_{1})
\ln\!\left(\frac{\eta_{1}}{\eta_{2}}\right)
 . \label{cprshn}
\end{equation}
The latter improves (\ref{unrshn}), whenever $\eta_{2}<\eta_{1}$.
The authors of \cite{cp2014} also derived entropic
uncertainty relations with quantum side information.

To the unitary matrix
$\wm=\bigl[\bigl[\langle{x}_{i}|z_{j}\rangle\bigr]\bigr]$, we assign
the set of all submatrices of class $k$, viz.
\begin{equation}
\mathcal{SUB}(\wm,k):=
\bigl\{
\mn\in\mset_{r\times{r}^{\prime}}(\zset):{\>}r+r^{\prime}=k+1,{\>}
\mn {\text{ is a submatrix of }} \wm
\bigr\}
\, . \label{subvk}
\end{equation}
The majorization relations are formulated in
terms of positive quantities
\begin{equation}
\zeta_{k}:=\max\bigl\{
\|\mn\|_{\infty}:{\>}\mn\in\mathcal{SUB}(\wm,k)
\bigr\}
\, . \label{skdf}
\end{equation}
Majorization relations of the tensor-product type
are formally posed as \cite{prz13}
\begin{equation}
p\otimes{q}\prec\omega^{\prime}
\, , \label{pq0w}
\end{equation}
where
\begin{equation}
\omega^{\prime}=(\xi_{1},\xi_{2}-\xi_{1},\ldots,\xi_{d}-\xi_{d-1})
\, ,
\qquad
\xi_{k}=\frac{(1+\zeta_{k})^{2}}{4}
\ . \label{wpdd}
\end{equation}
Majorization relations of the direct-sum type follow from the
formula \cite{rpz14}
\begin{equation}
p\oplus{q}\prec\{1\}\oplus\omega
\, , \label{pq1w}
\end{equation}
in which
$\omega=(\zeta_{1},\zeta_{2}-\zeta_{1},\ldots,\zeta_{d}-\zeta_{d-1})$
and $\zeta_{d}=1$ by the unitarity. The relations (\ref{pq0w}) and
(\ref{pq1w}) are both based on lemma 1 of \cite{prz13}. As
the right-hand side of its formula gives a correct upper bound for
any state, the majorization relations hold for all states.

It is very helpful to convert (\ref{pq0w}) and (\ref{pq1w}) into
inequalities between entropic functions. For $0<\alpha\neq1$, the
R\'{e}nyi $\alpha$-entropy is defined as \cite{renyi61}
\begin{equation}
R_{\alpha}(p):=\frac{1}{1-\alpha}\>
\ln\!\left(\sum\nolimits_{i}p_{i}^{\alpha}\right)
 . \label{repdf}
\end{equation}
It is not greater than the logarithm of the number of non-zero
probabilities. The Tsallis $\alpha$-entropy of degree
$0<\alpha\neq1$ reads as \cite{tsallis}
\begin{equation}
H_{\alpha}(p):=\frac{1}{1-\alpha}\>
\left(\sum\nolimits_{i} p_{i}^{\alpha}
- 1\right)
 . \label{tsent}
\end{equation}
In the limit $\alpha\to1$, both the entropies (\ref{repdf}) and
(\ref{tsent}) reduce to (\ref{shent}). For $\alpha\to+0$, the
entropy (\ref{repdf}) tends to the logarithm of the number of
non-zero probabilities. Basic properties of the above entropies with
applications are discussed in \cite{bengtsson}.

Since the R\'{e}nyi entropy is Schur concave, the majorization
relation (\ref{pq0w}) implies that, for $\alpha>0$, \cite{prz13}
\begin{equation}
R_{\alpha}(\clx;\bro)+R_{\alpha}(\clz;\bro)\geq{R}_{\alpha}(\omega^{\prime})
\, . \label{oldmr}
\end{equation}
The result (\ref{pq1w}) allows us to improve entropic
bounds \cite{rpz14}. For $0<\alpha\leq1$, one obtains
\begin{equation}
R_{\alpha}(\clx;\bro)+R_{\alpha}(\clz;\bro)\geq{R}_{\alpha}(\omega)
\, . \label{wnmr0}
\end{equation}
It is stronger, since $\omega\prec\omega^{\prime}$ and, therefore,
$R_{\alpha}(\omega)\geq{R}_{\alpha}(\omega^{\prime})$ \cite{rpz14}.
For $\alpha>1$, relation (\ref{wnmr0}) is not valid in general. The
authors of \cite{rpz14} proved another inequality
\begin{equation}
R_{\alpha}(\clx;\bro)+R_{\alpha}(\clz;\bro)\geq
\frac{2}{1-\alpha}{\>}
\ln\!\left(
\frac{1}{2}+\frac{1}{2}\,\sum\nolimits_{i}\omega_{i}^{\alpha}
\right)
 , \label{wnmr1}
\end{equation}
which holds for $\alpha>1$. The sum of two Tsallis
$\alpha$-entropies is bounded from below similarly to (\ref{wnmr0}).
For all $\alpha>0$, we have
\begin{equation}
H_{\alpha}(\clx;\bro)+H_{\alpha}(\clz;\bro)\geq{H}_{\alpha}(\omega)
\, . \label{wnmr01}
\end{equation}
Several generalizations of the above majorization relations were
also considered in the literature
\cite{prz13,rpz14,povmkz16,kraw18}.

The entropic framework allows one to take into account detection
inefficiencies, when the ``no-click'' event appears. To the given
detector efficiency $\vark\geq1/2$ and probability distribution
$\{p_{i}\}$, we assign a ``distorted'' distribution $p^{(\vark)}$
such that
\begin{equation}
p_{i}^{(\vark)}=\vark\, p_{i}
\, , \qquad
p_{\varnothing}^{(\vark)}=1-\vark
\, . \label{petad}
\end{equation}
The probability $p_{\varnothing}^{(\vark)}$ corresponds to the
no-click event. The above formulation was proposed in studying cycle
scenarios of the Bell type \cite{rchtf12}. It follows that
\cite{ramubs13,rastqic14}
\begin{equation}
H_{\alpha}\bigl(p^{(\vark)}\bigr)=\vark^{\alpha}H_{\alpha}(p)+h_{\alpha}(\vark)
\, , \label{qtlm0}
\end{equation}
where the binary entropy
$h_{\alpha}(\vark)=(1-\alpha)^{-1}\bigl(\vark^{\alpha}+(1-\vark)^{\alpha}-1\bigr)$.

\section{Uncertainty relations for flavor and mass}\label{sec3}

Let us proceed to uncertainty relations for measurements defined via
kets from the flavor basis
$\clf=\bigl\{|\nu_{e}\rangle,|\nu_{\mu}\rangle,|\nu_{\tau}\rangle\bigr\}$
and the mass basis
$\clm=\bigl\{|\nu_{1}\rangle,|\nu_{2}\rangle,|\nu_{3}\rangle\bigr\}$.
The flavor basis states are linked to the mass basis ones as
\begin{equation}
|\nu_{\beta}\rangle=
\sum\nolimits_{i}   u_{\beta{i}}^{*}\,|\nu_{i}\rangle
\, , \qquad
|\nu_{i}\rangle=
\sum\nolimits_{\beta} u_{\beta{i}}\,|\nu_{\beta}\rangle
\, . \label{intn}
\end{equation}
The $3\times{3}$-matrix $\um=[[u_{\beta{i}}]]$ is now referred to as
the Pontecorvo--Maki--Nakagawa--Sakata (PMNS) one
\cite{bruno58,mns62}. The PMNS matrix is further assumed to be
unitary, though this does not holds in some models. Questions of
testing such models are discussed in \cite{mohap07}. In the
case considered, the bases are orthonormal:
\begin{equation}
\langle\nu_{\beta}|\nu_{\gamma}\rangle=\delta_{\beta\gamma}
\, , \qquad
\langle\nu_{i}|\nu_{j}\rangle=\delta_{ij}
\, . \label{abij}
\end{equation}
The PMNS matrix is parametrized by mixing angles $\theta_{12}$,
$\theta_{23}$, $\theta_{13}$, and the CP-violating phase $\delta$,
viz.
\begin{align}
\um&=
\begin{pmatrix}
    1 & 0 & 0 \\
    0 & c_{23} & s_{23} \\
    0 & -s_{23} & c_{23}
\end{pmatrix}
\begin{pmatrix}
    c_{13} & 0 & s_{13}e^{-\iu\delta} \\
    0 & 1 & 0 \\
    -s_{13}e^{\iu\delta} & 0 & c_{13}
\end{pmatrix}
\begin{pmatrix}
    c_{12} & s_{12} & 0 \\
    -s_{12} & c_{12} & 0 \\
    0 & 0 & 1
\end{pmatrix}
\nonumber\\
&=
\begin{pmatrix}
   c_{12}c_{13} & s_{12}c_{13} & s_{13}e^{-\iu\delta} \\
   -s_{12}c_{23}-c_{12}s_{23}s_{13}e^{\iu\delta} & c_{12}c_{23}-s_{12}s_{23}s_{13}e^{\iu\delta} & s_{23}c_{13} \\
   s_{12}s_{23}-c_{12}c_{23}s_{13}e^{\iu\delta} & -c_{12}s_{23}-s_{12}c_{23}s_{13}e^{\iu\delta} & c_{23}c_{13}
\end{pmatrix}
 , \label{pmnsv}
\end{align}
where $c_{ij}\equiv\cos\theta_{ij}$ and
$s_{ij}\equiv\sin\theta_{ij}$. Since Majorana-like phases do not
affect majorization uncertainty relations, they are not recalled.

Within the considered framework, flavor and mass measurements are
formally treated in terms of the projectors
$|\nu_{\beta}\rangle\langle\nu_{\beta}|$ and
$|\nu_{i}\rangle\langle\nu_{i}|$, respectively. As flavor fields
contribute to the electroweak charged current, they can be produced
from the decay of a charged gauge boson \cite{schwindt20}. For
instance, neutrino--gauge boson interaction produces a lepton
further identified as electron, muon, or tau. In other words, flavor
can physically be measured through electroweak interactions.
Instead, the gravitation must inevitably participate in a mass
measurement. In practice, mass eigenstates can rather be observed by
absence of neutrino oscillations. Decay processes of charged gauge
bosons give a tool to entanglement-assisted determination of the
PMNS matrix \cite{schwindt20}. Of course, in neutrino physics any
question of experimental character is certainly not easy. Even so,
the measurements of interest are within capabilities of modern
physics. As will be discussed later, entropic formulation also
allows us to take into account detection inefficiencies.

\begin{table}
\caption{The NuFIT results based on data available in July 2020 \cite{nufit}.}
\begin{tabular}{@{}cccccccc@{}} \toprule
Without & $\sin^{2}\theta_{12}$ & $\theta_{12}$ in ${}^{\circ}$ & $\sin^{2}\theta_{23}$ & $\theta_{23}$ in ${}^{\circ}$ & $\sin^{2}\theta_{13}$ & $\theta_{13}$ in ${}^{\circ}$ & $\delta$ in ${}^{\circ}$ \\
\colrule
bfp $\pm1\bigma$ & $0.304_{-0.012}^{+0.013}$ & $33.44_{-0.75}^{+0.78}$ & $0.570_{-0.024}^{+0.018}$ & $49.0_{-1.4}^{+1.1}$ & $0.02221_{-0.00062}^{+0.00068}$ & $8.57_{-0.12}^{+0.13}$ & $195_{-25}^{+51}$ \\
$3\bigma$ range & $\,0.269\div0.343\,$ & $\,31.27\div35.86\,$ & $\,0.407\div0.618\,$ & $\,39.6\div51.8\,$ & $\,0.02034\div0.02430\,$ & $\,8.20\div8.97\,$ & $\,107\div403\,$ \\
\colrule
With SK & $\sin^{2}\theta_{12}$ & $\theta_{12}$ in ${}^{\circ}$ & $\sin^{2}\theta_{23}$ & $\theta_{23}$ in ${}^{\circ}$ & $\sin^{2}\theta_{13}$ & $\theta_{13}$ in ${}^{\circ}$ & $\delta$ in ${}^{\circ}$ \\
\colrule
bfp $\pm1\bigma$ & $0.304_{-0.012}^{+0.012}$ & $33.44_{-0.74}^{+0.77}$ & $0.573_{-0.020}^{+0.016}$ & $49.2_{-1.2}^{+0.9}$ & $0.02219_{-0.00063}^{+0.00062}$ & $8.57_{-0.12}^{+0.12}$ & $197_{-24}^{+27}$ \\
$3\bigma$ range & $\,0.269\div0.343\,$ & $\,31.27\div35.86\,$ & $\,0.415\div0.616\,$ & $\,40.1\div51.7\,$ & $\,0.02032\div0.02410\,$ & $\,8.20\div8.93\,$ & $\,120\div369\,$ \\
\botrule
\end{tabular}\label{tab1}
\end{table}

To concretize majorization uncertainty relations, we need to have
evaluate numbers $\zeta_{k}$ for the matrix (\ref{pmnsv}). Since
complete rows and columns of a unitary matrix are unit vectors, we
have $\zeta_{3}=1$. Roughly, the experimental findings are
consistent with a tri-bimaximal mixing matrix \cite{scott02}, which
for $\delta=0$ reads as
\begin{equation}
\begin{pmatrix}
   \sqrt{\frac{2}{3}} & \sqrt{\frac{1}{3}} & 0 \\
   -\sqrt{\frac{1}{6}} & \sqrt{\frac{1}{3}} & \sqrt{\frac{1}{2}} \\
   \sqrt{\frac{1}{6}} & -\sqrt{\frac{1}{3}}\, & \sqrt{\frac{1}{2}}
\end{pmatrix}
 . \label{trib}
\end{equation}
Due to (\ref{trib}), one has $\eta_{1}=\sqrt{2/3}$,
$\eta_{2}=\sqrt{1/2}$, and
$(\zeta_{1},\zeta_{2},\zeta_{3})=\bigl(\sqrt{2/3},1,1\bigr)$. These
values should be treated only as a first attempt. Let us quote some
results of NuFIT \cite{nufit,esgo20} with an updated global analysis
of neutrino oscillation measurements. We will use the best-fit
values based on data available in July 2020. The best-fit parameters
(bfp-values) from NuFIT \cite{nufit} are reproduced in Table
\ref{tab1}. The second of two imprints is obtained with the
inclusion of data on atmospheric neutrinos provided by
Super-Kamiokande \cite{skhome}. The $3\bigma$ ranges of two imprints
are shown as well.

Irrespectively to the inclusion of data on atmospheric neutrinos,
the term $\eta_{1}=\zeta_{1}$ is equal to $|u_{e1}|=c_{12}c_{13}$
within the $3\bigma$ ranges provided by NuFIT \cite{nufit}. Under
the same circumstances, we have
$\eta_{2}=c_{13}\max\{c_{23},s_{23}\}$. Appositely, this choice of
$\eta_{1}$ and $\eta_{2}$ remains valid with elements of the matrix
(\ref{trib}). With the bfp-values, the right-hand side of
(\ref{cprshn}) exceeds the Maassen--Uffink bound $-2\ln\eta_{1}$
 by $4.5$ \%. To write explicitly majorization uncertainty
relations, more narrow ranges should be taken. It turned out that
$\zeta_{2}=c_{13}$ within the $1\bigma$ ranges given by NuFIT
\cite{nufit}. This conclusion is also independent of the inclusion
of data on atmospheric neutrinos. With the bfp-values and
$\alpha=1$, the right-hand side of (\ref{wnmr0}) exceeds
$-2\ln\eta_{1}$ by around $31.3$ \%. The structure of (\ref{pmnsv})
is such that
\begin{equation}
\sqrt{|u_{e1}|^{2}+|u_{e2}|^{2}}=c_{13}=
\sqrt{|u_{\mu3}|^{2}+|u_{\tau3}|^{2}}
\, . \label{mue12}
\end{equation}
It is natural that the answer $\zeta_{2}=c_{13}$ also follows from
(\ref{trib}). From the unitarity condition, we inevitably have
$\zeta_{3}=1$. In contrast to the NuFIT results, the matrix
(\ref{trib}) leads to $\zeta_{2}=1$. In each of the relations
(\ref{oldmr}), (\ref{wnmr0}) and (\ref{wnmr1}), the right-hand side
then becomes the binary entropy. It is instructive to visualize this
distinction.

\begin{figure}
\centering \includegraphics[height=6.1cm]{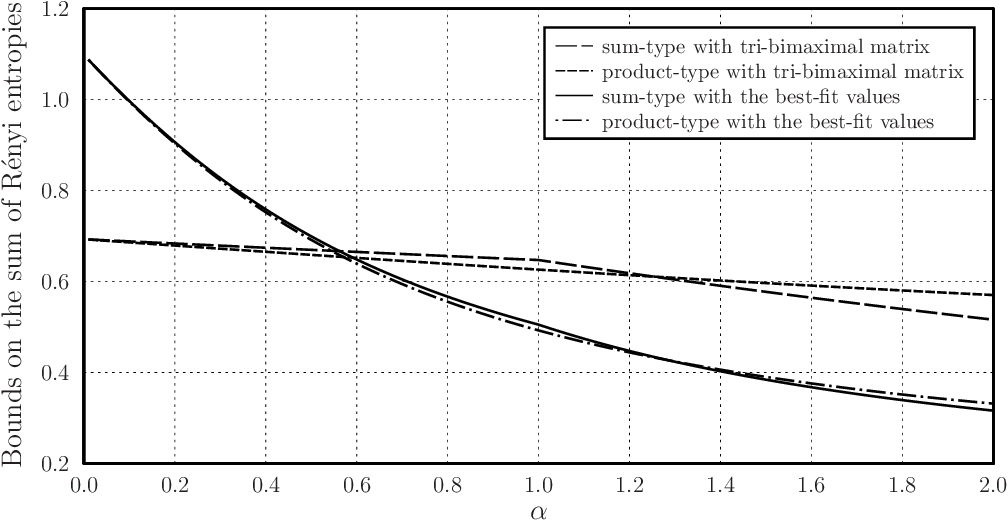}
\caption{\label{fg1} Comparing the product-type lower bound (\ref{oldmr}) and the sum-type one given by (\ref{wnmr0}) and (\ref{wnmr1}) for $\alpha$ between $0$ and $2$.}
\end{figure}

The product-type inequality (\ref{oldmr}) holds for all
$\alpha>0$. Another entropic inequality is expressed as
(\ref{wnmr0}) for $0<\alpha\leq1$ and as (\ref{wnmr1}) for
$\alpha>1$. For $0<\alpha\leq2$, the right-hand sides of these
inequalities are shown in Fig. \ref{fg1} with the use of the NuFIT
best-fit values and the elements of (\ref{trib}).
For concreteness, the corresponding two curves use the best-fit
values with the inclusion of data on atmospheric neutrinos
\cite{nufit}. A contrast with the best-fit values without such data
is so small that does not have actual bearing. In
opposite, substituting the elements of (\ref{trib}) leads to
principally different curves. It is natural since the use of
(\ref{trib}) gives $s_{13}=0$ and, further,
$\omega_{3}=\omega^{\prime}_{3}=0$. So, the entropies
$R_{\alpha}(\omega)$ and $R_{\alpha}(\omega^{\prime})$ become
binary. For $\alpha\to+0$, the corresponding two curves approach
$\ln2$ instead of $\ln3$. Probing the number of non-zero
probabilities, R\'{e}nyi entropies with small $\alpha$ mirror the
number of neutrino species. The curves of Fig. \ref{fg1} witness a
sensitivity of the majorization bounds to $\theta_{13}$, especially
with varying $\alpha$. This gives an example of utility of
parametrized uncertainty relations \cite{maass}. Of course, these
curves also reflect some features of deriving the used uncertainty
bounds. Nonetheless, relations with parametrized entropies quite
deserve to be used, whenever we become delving into the flavor-mass
complementarity.

Let obtained data with respect to the bases $\clf$ and $\clm$ be
characterized by the efficiencies $\vark_{\clf}$ and $\vark_{\clm}$,
respectively. Combining (\ref{wnmr01}) with (\ref{qtlm0}) gives
\begin{equation}
H_{\alpha}^{(\vark_{\clf})}(\clf;\bro)+H_{\alpha}^{(\vark_{\clm})}(\clm;\bro)
\geq\vark^{\alpha}H_{\alpha}(\omega)+h_{\alpha}(\vark_{\clf})+h_{\alpha}(\vark_{\clm})
\, . \label{alvark}
\end{equation}
where $\vark=\min\{\vark_{\clf},\vark_{\clm}\}\geq1/2$.
By $H_{\alpha}^{(\vark_{\clf})}(\clf;\bro)$ and
$H_{\alpha}^{(\vark_{\clm})}(\clm;\bro)$, we mean the
$\alpha$-entropies calculated with ``distorted'' actual
distributions. In particular, the Shannon entropies obey
\begin{equation}
H_{1}^{(\vark_{\clf})}(\clf;\bro)+H_{1}^{(\vark_{\clm})}(\clm;\bro)
\geq\vark\,H_{1}(\omega)+h_{1}(\vark_{\clf})+h_{1}(\vark_{\clm})
\, . \label{shvark}
\end{equation}
Replacing $H_{1}(\omega)$ with $-2\ln\eta_{1}$ here, we get the
corresponding corollary of the the Maassen--Uffink relation
\cite{maass}. Thus, detection inefficiencies produce additional
uncertainties in the entropies of actually measured data. The above
discussion completes the results of \cite{schwindt20} in
this regard as well. As follows from the performed inspection, only
$\theta_{12}$ and $\theta_{13}$ are actually needed to apply
(\ref{pq0w}) and (\ref{pq1w}) in the case of interest. Hence, the
inequality (\ref{shvark}) is valid within the same framework as
corollaries of the Maassen--Uffink uncertainty relation.

\section{On uncertainty relations with quantum side information}\label{sec4}

The authors of \cite{schwindt20} formulated a protocol to
probe parameters of the PMNS matrix from quantum manipulations and
measurements on entangled neutrino-lepton pairs. For such
experiments, the ``quantum-memory'' uncertainty relation of 
\cite{BCCRR10} was proposed to be used. As was already mentioned,
one very likely deals with the case, when $\eta_{2}$ is strictly
less than $\eta_{1}$. Hence, the stronger results of 
\cite{cp2014} should also be kept in mind. Entangled
neutrino-lepton pairs could result from decay processes of the type
$W^{+}\mapsto\nu+\ell^{+}$, namely
\begin{equation}
\frac{1}{\sqrt{3}}\,\bigl(|\nu_{e}\rangle\otimes|e^{+}\rangle+|\nu_{\mu}\rangle\otimes|\mu^{+}\rangle+|\nu_{\tau}\rangle\otimes|\tau^{+}\rangle\bigr)=
\frac{1}{\sqrt{3}}\,\sum\nolimits_{j}|\nu_{j}\rangle\otimes|\tilde{\ell}_{j}^{+}\rangle
\, . \label{enln}
\end{equation}
As direct manipulations in flavor space are challenging, projective
measurements in the basis of leptonic states
$|\tilde{\ell}_{j}^{+}\rangle$ can be used instead
\cite{schwindt20}. In each summand of the right-hand side of
(\ref{enln}), the corresponding neutrino is in a definite mass
eigenstate and does not show oscillations. To implement proper
manipulations on the lepton side, the exact matrix elements are
required. In reality, however, actual values of the four angles
should be treated only as a guess. This motivates the use of
uncertainty relations with quantum memory.

Uncertainty relations in the presence of quantum memory are
posed as follows \cite{BCCRR10,cp2014}. Let $\bro_{AB}$ be density
matrix of a system of two subsystems $A$ and $B$. The reduced
densities are obtained by partial tracing, viz.
\begin{equation}
\bro_{A}=\Tr_{B}(\bro_{AB})
\, , \qquad
\bro_{B}=\Tr_{A}(\bro_{AB})
\, . \label{broab}
\end{equation}
To the given orthonormal basis $\clx=\bigl\{|x_{i}\rangle\bigr\}$ in
$\hh_{A}$, we assign the linear map
\begin{equation}
\bro_{A}\mapsto
\Phi_{\clx}(\bro_{A}):=
\sum\nolimits_{i}|x_{i}\rangle\langle{x}_{i}|\bro_{A}|x_{i}\rangle\langle{x}_{i}|
\, . \label{phiax}
\end{equation}
Taking the two bases $\clx=\bigl\{|x_{i}\rangle\bigr\}$ and
$\clz=\bigl\{|z_{j}\rangle\bigr\}$ in $\hh_{A}$, we further define
the density matrices
\begin{equation}
\bro_{XB}=(\Phi_{\clx}\otimes\id)(\bro_{AB})
\, , \qquad
\bro_{ZB}=(\Phi_{\clz}\otimes\id)(\bro_{AB})
\, , \label{brxz}
\end{equation}
where $\id:\>\bro_{B}\mapsto\bro_{B}$ is the identity map. The
quantum conditional entropy $S_{1}(A|B)$ is defined as
\cite{nielsen}
\begin{equation}
S_{1}(A|B):=S_{1}(\bro_{AB})-S_{1}(\bro_{B})
\, , \label{qonde}
\end{equation}
where $S_{1}(\vbro)=-\,\Tr(\vbro\ln\vbro)$ denotes the von Neumann
entropy. Coles and Piani showed that \cite{cp2014}
\begin{equation}
S_{1}(X|B)+S_{1}(Z|B)\geq-2\ln\eta_{1}+(1-\eta_{1})
\ln\!\left(\frac{\eta_{1}}{\eta_{2}}\right)+S_{1}(A|B)
\, . \label{copi14}
\end{equation}
The two terms $S_{1}(X|B)$ and $S_{1}(Z|B)$ are nonnegative as
entropies of classical probability distributions. On the other hand,
the conditional entropy $S_{1}(A|B)$ can be negative if $\bro_{AB}$
is entangled. For pure states of the form (\ref{enln}), we have
$S_{1}(\bro_{AB})=0$ and $S_{1}(\bro_{A})=\ln3$. The second term in
the right-hand side of (\ref{copi14}) is nonnegative. Replacing it
with zero, the uncertainty relation of \cite{BCCRR10}
follows,
\begin{equation}
S_{1}(X|B)+S_{1}(Z|B)\geq-2\ln\eta_{1}+S_{1}(A|B)
\, . \label{bccrr}
\end{equation}
The latter is mentioned \cite{schwindt20} as a tool for
entanglement-assisted determination of the PMNS matrix. The result
(\ref{copi14}) is also suitable to use in this context. In the case
of product states, we have the additivity property \cite{nielsen}
\begin{equation}
S_{1}\bigl(\bro_{A}\otimes\bro_{B}\bigr)=S_{1}(\bro_{A})+S_{1}(\bro_{B})
\, . \label{ditiv}
\end{equation}
If the measured system is not coupled with others, the
inequality (\ref{copi14}) reduces to
\begin{equation}
H_{1}(\clx;\bro_{A})+H_{1}(\clz;\bro_{A})\geq-2\ln\eta_{1}+(1-\eta_{1})
\ln\!\left(\frac{\eta_{1}}{\eta_{2}}\right)+S_{1}(\bro_{A})
\, . \label{copi14a}
\end{equation}
Without $S_{1}(\bro_{A})\geq0$, the latter gives (\ref{cprshn}).
Vanishing the second term in the right-hand side of (\ref{copi14a}),
we get the Maassen--Uffink bound added by the von Neumann entropy of
the measured state, namely
\begin{equation}
H_{1}(\clx;\bro_{A})+H_{1}(\clz;\bro_{A})\geq-2\ln\eta_{1}+S_{1}(\bro_{A})
\, . \label{copi14b}
\end{equation}
As was shown in \cite{ccyz12}, this result also follows from
the monotonicity of the quantum relative entropy. Thus, the relation
(\ref{copi14a}) should be used together with (\ref{copi14b}). But the
former needs concrete values of all three mixing angles.

\section{Conclusions}\label{sec5}

We examined applications of majorization uncertainty relations to
neutrino states in the three-dimensional flavor-mass space. The
recent literature has brought an attention to situations in which
flavor-mass uncertainty relations could be used. Since neutrinos are
almost free from decoherence effects, they are candidates to carry
quantum information. In practice, of course, their applicability in
quantum technology is an open question. Applications of
information-theoretic concepts deserve to be involved in experiments
to analyze the cosmic neutrino background. The authors of 
\cite{schwindt20} proposed an approach with entangled states to
probe experimentally parameters on the mixing matrix. The amount of
accumulated experimental data allows us to made the following
observations.

Applying the entropic approach to flavor-mass uncertainties is
coupled with incorporating concrete values of the mixing angles. In
effect, the Maassen-Uffink relation as well as the majorization
approach use only two concrete parameters, namely $\theta_{12}$ and
$\theta_{13}$. At first glance, majorization uncertainty relations
may seem to require all the matrix elements to be known. One of
unexpected results of our research is that only $\theta_{12}$ and
$\theta_{13}$ are actually needed, at least within the current level
of experience concerning the PMNS matrix. Hence, the majorization
approach is valid within the same framework as the Maassen--Uffink
relation discussed for the flavor-mass pair in 
\cite{schwindt20}. In this way, lower entropic bounds are
improved by almost one third.

So, the use of two mixing angles characterizes the two of three
considered schemes to pose desired entropic uncertainty relations.
Further, all the three mixing angles should be put to use the
uncertainty relation of Coles and Piani \cite{cp2014}. The same
conclusion holds, when the results of \cite{cp2014} are
assumed to be used in entanglement-assisted tests of the mixing
matrix. The case of detection inefficiencies is naturally
incorporated into the entropic framework. Such inefficiencies will
produce some additional level of uncertainty. Thus, majorization
uncertainty relations deserve to be used in this context among other
formulations. The presented results may be appled in future studies
of neutrinos.


\begin{thebibliography}{000}


\bibitem{heisenberg}
W.~Heisenberg, {\it Z. Phys.} {\bf 43}, 172 (1927).

\bibitem{kennard}
E.H.~Kennard, {\it Z. Phys.} {\bf 44}, 326 (1927).

\bibitem{robert}
H.P.~Robertson, {\it Phys. Rev.} {\bf 34}, 163 (1929).

\bibitem{deutsch}
D.~Deutsch, {\it Phys. Rev. Lett.} {\bf 50}, 631 (1983).

\bibitem{maass}
H.~Maassen and J.B.M.~Uffink, {\it Phys. Rev. Lett.} {\bf 60}, 1103 (1988).

\bibitem{hirs}
I.I.~Hirschman, {\it Amer. J. Math.} {\bf 79}, 152 (1957).

\bibitem{beck}
W.~Beckner, {\it Ann. Math.} {\bf 102}, 159 (1975).

\bibitem{birula1}
I.~Bia{\l}ynicki-Birula and J.~Mycielski, {\it Commun. Math. Phys.} {\bf 44}, 129 (1975).

\bibitem{ww10}
S.~Wehner and A.~Winter, {\it New J. Phys.} {\bf 12}, 025009 (2010).

\bibitem{brud11}
I.~Bia{\l}ynicki-Birula and {\L}.~Rudnicki, Entropic uncertainty
relations in quantum physics, in {\it Statistical Complexity}, ed. K.D. Sen (Springer, 2011), pp. 1--34.

\bibitem{cbtw17}
P.J.~Coles, M.~Berta, M.~Tomamichel, and S.~Wehner, {\it Rev. Mod. Phys.} {\bf 89}, 015002 (2017).

\bibitem{cerf19}
A.~Hertz and N.J.~Cerf, {\it J. Phys. A: Math. Theor.} {\bf 52}, 173001 (2019).

\bibitem{BCCRR10}
M.~Berta, M.~Christandl, R.~Colbeck, J.M.~Renes, and R.~Renner, {\it Nature Phys.} {\bf 6}, 659 (2010).

\bibitem{kraus87}
K.~Kraus, {\it Phys. Rev. D} {\bf 35}, 3070 (1987).

\bibitem{riesz27}
M.~Riesz, {\it Acta Math.} {\bf 49}, 465 (1927).

\bibitem{rans12}
A.E.~Rastegin, {\it Int. J. Theor. Phys.} {\bf 21}, 1300 (2012).

\bibitem{zbp13}
S.~Zozor, G.M.~Bosyk, and M.~Portesi, {\it J. Phys. A: Math. Theor.} {\bf 46}, 465301 (2013).

\bibitem{zbp14}
S.~Zozor, G.M.~Bosyk, and M.~Portesi, {\it J. Phys. A: Math. Theor.} {\bf 47}, 495302 (2014).

\bibitem{prtv11}
M.H.~Partovi, {\it Phys. Rev. A} {\bf 84}, 052117 (2011).

\bibitem{prz13}
Z.~Pucha{\l}a, {\L}.~Rudnicki, and K.~\.{Z}yczkowski, {\it J. Phys. A: Math. Theor.} {\bf 46}, 272002 (2013).

\bibitem{fgg13}
S.~Friedland, V.~Gheorghiu, and G.~Gour, {\it Phys. Rev. Lett.} {\bf 111}, 230401 (2013).

\bibitem{rpz14}
{\L}.~Rudnicki, Z.~Pucha{\l}a, and K.~\.{Z}yczkowski, {\it Phys. Rev. A} {\bf 89}, 052115 (2014).

\bibitem{zuber12}
K.~Zuber, {\it Neutrino Physics} (CRC Press, 2012).

\bibitem{bruno57}
B.~Pontecorvo, {\it Sov. Phys. JETP} {\bf 6}, 429 (1957).

\bibitem{kajita16}
T.~Kajita, {\it Ann. Phys. (Berlin)} {\bf 528}, 459 (2016).

\bibitem{mcdd16}
A.B.~McDonald, {\it Ann. Phys. (Berlin)} {\bf 528}, 469 (2016).

\bibitem{bruno58}
B.~Pontecorvo, {\it Sov. Phys. JETP} {\bf 7}, 172 (1958).

\bibitem{mns62}
Z.~Maki, M.~Nakagawa, and S.~Sakata, {\it Prog. Theor. Phys.} {\bf 28}, 870 (1962).

\bibitem{schwindt20}
S.~Floerchinger and J.-M.~Schwindt, {\it Phys. Rev. D} {\bf 102}, 093001 (2020).

\bibitem{cp2014}
P.J.~Coles and M.~Piani, {\it Phys. Rev. A} {\bf 89}, 022112 (2014).

\bibitem{hj1990}
R.A.~Horn and C.R.~Johnson, {\it Matrix Analysis} (Cambridge University Press, 1990).

\bibitem{renyi61}
A.~R\'{e}nyi, On measures of entropy and information, in {\it Proc.
4th Berkeley Symposium on Mathematical Statistics and Probability}, ed. J. Neyman
 (University of California Press, 1961), pp. 547--561.

\bibitem{tsallis}
C.~Tsallis, {\it J. Stat. Phys.} {\bf 52}, 479 (1988).

\bibitem{bengtsson}
I.~Bengtsson and K.~\.{Z}yczkowski, {\it Geometry of Quantum States: An Introduction to Quantum Entanglement} (Cambridge University Press, 2017).

\bibitem{povmkz16}
A.E.~Rastegin and K.~\.{Z}yczkowski, {\it J. Phys. A: Math. Theor.} {\bf 49}, 355301 (2016).

\bibitem{kraw18}
Z.~Pucha{\l}a, {\L}.~Rudnicki, A.~Krawiec, and K.~\.{Z}yczkowski, {\it J. Phys. A: Math. Theor.} {\bf 51}, 175306 (2018).

\bibitem{rchtf12}
R.~Chaves and T.~Fritz, {\it Phys. Rev. A} {\bf 85}, 032113 (2012).

\bibitem{ramubs13}
A.E.~Rastegin, {\it Eur. Phys. J. D} {\bf 67}, 269 (2013).

\bibitem{rastqic14}
A.E.~Rastegin, {\it Quantum Inf. Comput.} {\bf 14}, 0996 (2014).

\bibitem{mohap07}
R.N.~Mohapatra, S.~Antusch, K.S.~Babu, {\it et al.}, {\it Rep. Prog. Phys.} {\bf 70}, 1757 (2007).

\bibitem{scott02}
P.F.~Harrison, D.H.~Perkins, and W.G.~Scott, {\it Phys. Lett. B} {\bf 530}, 167 (2002).

\bibitem{nufit}
http://www.nu-fit.org

\bibitem{esgo20}
I.~Esteban, M.C.~Gonzalez-Garcia, M.~Maltoni, I.~Martinez-Soler, T.~Schwetz, and A.~Zhou, {\it JHEP} {\bf 09}, 178 (2020).

\bibitem{skhome}
http://www-sk.icrr.u-tokyo.ac.jp/sk/index-e.html

\bibitem{nielsen}
M.A.~Nielsen and I.L.~Chuang, {\it Quantum Computation and Quantum Information} (Cambridge University Press, 2000).

\bibitem{ccyz12}
P.J.~Coles, R.~Colbeck, L.~Yu, and M.~Zwolak, {\it Phys. Rev. Lett.} {\bf 108}, 210405 (2012).

\end{thebibliography}
\end{document}